\documentclass[lettersize,journal]{IEEEtran}

\usepackage{cite}                    
\usepackage{amsmath,amssymb,amsfonts}
\usepackage[numbers,sort&compress]{natbib}
\usepackage{amsthm}                  
\usepackage{mathtools}               
\usepackage{graphicx}                
\usepackage{subcaption}              
\usepackage{xcolor}                  
\usepackage{textcomp}                
\usepackage{booktabs}                
\usepackage{multirow}                
\usepackage{array}                   
\usepackage{threeparttable}          
\usepackage{enumitem}                
\usepackage{algorithm}               
\usepackage[noend]{algpseudocode}    
\usepackage{listings}                
\usepackage{tcolorbox}               
\tcbuselibrary{breakable,skins}
\usepackage{fontawesome}             
\usepackage{url}                     
\usepackage[normalem]{ulem}          
\useunder{\uline}{\ul}{}

\newcommand{\TECH}{\textit{AssertMate}}

\newcommand{\ChatAssert}{\textit{ChatAssert}{$^\ast$}}

\newcommand{\EditAS}{\textit{EditAS}}
\newcommand{\TOGLL}{\textit{TOGLL}}
\newcommand{\RetriGen}{\textit{RetriGen}}
\newcommand{\AVCAgent}{ActVCon}
\newcommand{\MAGE}{EVGen}
\newcommand{\EVGenerator}{EVGen}
\newcommand{\ins}[1]{{\color{blue}#1}}

\newcommand{\Design}{\noindent \uline{\textbf{Process:}}}
\newcommand{\ResultAndAnalysis}{\noindent \uline{\textbf{Results:}}}
\definecolor{awesome}{rgb}{0.0, 0.2, 0.6}

\newtcolorbox{summarybox}[1]{
    colback=cyan!5!white,
    colframe=blue!50!cyan!80!white, 
    title={#1},
    coltitle=white,          
    fonttitle=\bfseries,    
    boxrule=0pt,
    arc=3pt,
    shadow={2pt}{-2pt}{gray!25}
}

\definecolor{backcolour}{rgb}{0.95,0.95,0.92}

\definecolor{commentorange}{RGB}{180,90,0}

\begin{document}

\title{Agent-Based Test Assertion Generation via Diverse Perspective Aggregation}

\author{
Dong~Wang, 
Qiaoyu~Han,
Lin~Yang,  
Jianyi~Zhou, 
Guangtai Liang,
and~Junjie~Chen*%
\IEEEcompsocitemizethanks{
\IEEEcompsocthanksitem Dong Wang, Qiaoyu Han, Lin Yang, and Junjie Chen are with the College of Intelligence and Computing, Tianjin University, Tianjin, China.
\protect\\ E-mail: \{dong\_w, hanqiaoyu, linyang, junjiechen\}@tju.edu.cn
\IEEEcompsocthanksitem Jianyi Zhou and Guangtai Liang are with Huawei Cloud Computing Technologies Co., Ltd., Beijing, China.
E-mail: \{zhoujianyi2, liangguangtai\}@huawei.com
\IEEEcompsocthanksitem Corresponding author: Junjie Chen.
}
}

\markboth{IEEE Transactions on Software Engineering}%
{Wang \MakeLowercase{\textit{et al.}}: Agent-Based Test Assertion Generation via Diverse Perspective Aggregation}

\IEEEtitleabstractindextext{
\begin{abstract}
Test assertions are critical elements of unit tests, serving as checkpoints to validate expected behavior and ensure software correctness. 
Numerous techniques have been proposed to automate assertion generation, with recent progress notably driven by large language models (LLMs).
Despite the promise, existing approaches such as ChatAssert suffer from modest accuracy, heavy reliance on oversampling, and vulnerability to model randomness due to one-shot prompting.
To address these limitations, we propose \TECH{}, a novel agent-based assertion generation framework that enhances the quality and reliability of LLM-generated assertions through three key components: (1) actual value construction that identifies assertion targets via static analysis and type-aware heuristics; (2) multi-perspective expected value prediction using code generation, retrieval-augmented generation (RAG), and chain-of-thought (CoT) reasoning agents; 
and (3) an LLM-as-a-Judge collaboration mechanism to select the most appropriate assertion.
Evaluation on the Defects4J benchmark demonstrates that \TECH{} significantly outperforms state-of-the-art techniques in compilation success and pass rates, along with substantially higher bug detection capabilities.
Integration with EvoSuite further validates \TECH{}'s practicality, yielding superior mutation coverage and kill counts.
Ablation studies reveal that each of the three components makes a significant and complementary contribution to the overall performance. 
This work affirms the great potential of aggregating diverse perspectives to enhance the effectiveness of LLM-based assertion generation.
\end{abstract}

}

\maketitle
\IEEEdisplaynontitleabstractindextext
\IEEEpeerreviewmaketitle


\section{Introduction}
\label{sec:introduction}
\IEEEPARstart{U}{nit} testing plays a pivotal role in software quality assurance by verifying functional correctness and ensuring that implementations align with specified requirements~\cite{ChatAssert,ChatTester,ChatUnitest,EditAS,UTLLMStudy,ATLAS,IRAG,TOGLL,Evosuite,Randoop}.
A complete unit test consists of two essential components: a test prefix, which constructs the execution context and exercises the focal method, and one or more assertions, which serve as test oracles by specifying the expected behavior.
While recent progress in automated LLM-based test generation has substantially improved the generation of executable test prefixes, constructing correct and effective assertions remains a fundamental challenge.
This is because assertions ultimately determine whether the observed program behavior conforms to its expected behavior; consequently, even tests with high code coverage may fail to reveal defects when their assertions are weak, incomplete, or semantically incorrect, further limiting their fault-detection capability in practice.
Therefore, test assertion generation has become an important standalone research problem, focusing on the construction of high-quality test oracles regardless of how test prefixes are obtained.

To automate the assertion generation, prior research has introduced various types of automated assertion generation techniques, including traditional techniques such as Symbolic Execution~\cite{DBLP:conf/relsoft/BoyerEL75,DBLP:journals/tse/Howden77,DBLP:conf/ibm/King74,DBLP:conf/tacas/KhurshidPV03,DBLP:conf/osdi/CadarDE08,DBLP:conf/kbse/PasareanuR10} and Specification Mining~\cite{DBLP:conf/kbse/ZhengMLXK11,DBLP:conf/icst/TanMTL12,DBLP:conf/issta/GoffiGEP16}.
While being effective in certain contexts, these traditional techniques suffer from notable limitations, particularly the path explosion~\cite{DBLP:journals/csur/BaldoniCDDF18} and the lack of structured documentation in many real-world projects~\cite{TOGA}.
Recent advancements in deep learning (DL) and large language models (LLMs) have significantly improved automated assertion generation~\cite{ATLAS,TOGLL,IRAG,EditAS,ChatAssert}. By leveraging their strong code comprehension capabilities, these models often outperform traditional techniques. 
Typically, these approaches take as input a focal method (i.e., the method under test) and a test prefix (i.e., a test body without assertions), and use either DL models or LLMs to generate assertions.
Among them, ChatAssert~\cite{ChatAssert} stands as the state-of-the-art LLM-based assertion generation technique, utilizing the power of ChatGPT.
Nevertheless, while ChatAssert advances beyond earlier techniques, it still faces critical limitations and presents opportunities for further improvement.
Its top-1 accuracy remains modest at 0.45, with only a slight increase to 0.54 in top-10 accuracy, even after applying multi-round post-processing to iteratively fix compilation and runtime errors.
This first highlights substantial uncertainty in generating accurate assertions on the initial attempt.
Moreover, ChatAssert's heavy reliance on over-sampling (generating up to 10 assertions to cover a likely correct one) exposes the inefficiency and search space under-exploration of single-prompt inference, which also increases the burden of manual validation in practice.
The 9-percentage-point gain further reflects its susceptibility to LLM randomness, limiting its reliability in contexts that demand deterministic and precise test oracles.

These limitations align with findings from recent work~\cite{self-construct}, which observed that one-shot LLM queries often yield overconfident or erratic outputs, resulting in stubborn or inconsistent feedback during self-evaluation. 
The study further suggests that incorporating diverse problem-solving perspectives provides a more robust foundation for iterative refinement and can significantly improve output quality. 
Inspired by this insight, we theorize that adopting a multi-perspective prompting strategy holds promise for enhancing the effectiveness and reliability of LLM-based assertion generation.
Meanwhile, it brings several unique challenges in this context.
\textit{First}, due to the under-specified nature of unit tests, the method for verifying the focal methods' intended behavior is often ambiguous and lacks explicit contextual cues.
LLMs may randomly interpret the test target differently per perspective-based query, leading to inconsistent or inaccurate assertions.  
\textit{Second}, a multi-prompting strategy relies on selecting and combining diverse reasoning approaches, with each agent potentially emphasizing different aspects of the code. 
An inappropriate choice or combination of strategies may introduce noise or 
redundancy, undermining the complementarity from multiple perspectives, thus reducing the quality of generated assertions.
\textit{Third}, when multiple agents generate candidate assertions, selecting the most accurate and semantically aligned one becomes a non-trivial task.

In this paper, we propose \textbf{\TECH{}}, a novel agent-based assertion generation framework that effectively leverages LLMs for producing high-quality test assertions.
\TECH{} is designed with three core components, each tailored to address key limitations of prior work and challenges introduced by multi-perspective prompting:
\begin{itemize}[leftmargin=10pt]
    \item \textbf{Accurate Identification of Assertion Targets.} 
    Given a focal method and its test prefix, \TECH{} first performs static analysis, assisted by LLMs, to identify a set of test targets, such as return values and modified attributes. 
    It then constructs the corresponding actual values (assertion targets) using a rule-based and type-aware strategy. 
    This process effectively clarifies ``what to test'' and narrows the task to expected value prediction, thereby reducing ambiguity and providing a solid foundation for the subsequent multi-perspective assertion generation.
    \item \textbf{Expected Value Generation from Diverse Perspectives.}
    To address the overconfidence and bias inherent in single-prompt strategies, \TECH{} adopts a multi-perspective reasoning approach for expected-value prediction (i.e., inferring the desired outcomes corresponding to the constructed actual values).
    Here, a perspective refers to a distinct reasoning strategy characterized by its information source and inference mechanism.
    Specifically, \TECH{} employs three complementary perspectives: code-generation-based reasoning, which infers expected values directly from program semantics; retrieval-augmented reasoning (RAG), which grounds prediction on semantically similar examples; and chain-of-thought (CoT) reasoning, which derives expected values through explicit intermediate reasoning.
    

    To further improve reliability, \TECH{} aggregates the candidate assertions generated from these complementary perspectives through a probability-based reranking mechanism that prioritizes the most promising candidates, thereby reducing the search space. The remaining candidates are then passed to the subsequent collaboration stage for final selection, effectively answering the question of ``what to expect.''

    \item \textbf{``LLM-as-a-Judge'' Collaboration Mechanism.} 
    To enhance decision-making capabilities for selecting high-quality assertions, \TECH{} introduces a Judge agent based on the ``LLM-as-a-Judge''~\cite{LLMAsJudge} paradigm. 
    This agent conducts an in-depth comparative analysis of candidate assertions, leveraging a CoT reasoning strategy to evaluate their semantic correctness, coherence with the test context, and alignment with the actual values, ultimately selecting the most appropriate assertion.
\end{itemize}
To summarize, \TECH{}’s novelty lies in a structured shift from monolithic assertion generation to a decomposed and software-engineered workflow, built upon two key design contributions:

(I) Task Decoupling with Domain-Specific Optimization.
By decomposing assertion generation into structured subtasks, such as actual value derivation through program analysis, \TECH{} reduces LLM hallucinations and improves reliability.

(II) Synergistic Agent Design.
\TECH{} combines complementary perspectives, actual-value construction, and expected-value prediction through modular components. It first uses confidence-based reranking to select one candidate from each perspective and then uses the Judge agent for cross-perspective selection, addressing the ``last-mile'' challenge without manual top-$k$ inspection.

We evaluated \TECH{} on the Defects4J~\cite{DBLP:conf/issta/JustJE14} benchmark and compared its effectiveness against four widely adopted or state-of-the-art baseline techniques across both DL-based and language model-based approaches: \EditAS{}, \TOGLL{}, \textit{RetriGen}, and \ChatAssert{} (without its repair module).
Across five independent runs, \TECH{} achieves mean Compilation Success Rate (CSR) and Pass Rate (PR) values of 76.78\% and 60.09\%, respectively; the corresponding sample standard deviations (SDs) are 0.90 and 1.07 percentage points. These results outperform the strongest baseline, \ChatAssert{}, by 52.92\% and 50.94\% in relative terms, respectively.
Its mean Bug Detection Rate (BDR) is 46.25\%, and the corresponding SD is 1.75 percentage points. In comparison, \ChatAssert{} achieves a mean BDR of 37.32\% with an SD of 1.82 percentage points. After averaging the five per-bug outcomes, a paired Wilcoxon signed-rank test indicates a significant improvement over \ChatAssert{} ($p=0.0489$, $delta=0.152$).

Furthermore, we examined the practical utility of integrating \TECH{} with Evosuite~\cite{Evosuite}. 
Results show that the assertions generated by \TECH{} consistently yield higher CSRs and achieve the highest number of covered and killed mutants among all compared techniques. Its killed-mutant count improves by 40.49\%--492.76\% over \ChatAssert{}, \TOGLL{}, and \EditAS{}.
Ablation studies further validate the substantial contribution of \TECH{}’s three core components: actual value constructor, multi-perspective expected value generation, and the collaboration mechanism. 
Notably, the actual value constructor not only mitigates hallucinations and guides LLM reasoning during assertion generation but also enhances existing LLM-based techniques when incorporated.

\smallskip
\noindent
\textbf{Contributions.} The key contributions of this paper are:  
\begin{itemize}[leftmargin=10pt]
    \item We propose a novel agent-based assertion generation framework to enhance assertion quality, with a full replication package publicly released for reproducibility~\cite{homepage}.  
    \item Experimental results comprehensively demonstrate that \TECH{} excels at generating syntactically correct and semantically rich assertions, consistently delivering superior bug-detection performance compared to all baselines.
    \item Evaluation of \TECH{}’s integration with EvoSuite affirms its practicality and scalability in real-world testing scenarios.
\end{itemize}


\section{Background and Related Work}
\label{sec:terminology}
In this section, we introduce the concept of fine-grained assertion terminology and review related work on automated test assertion generation techniques.

\subsection{Fine-Grained Assertion Terminology}
\begin{figure}
    \centering
    \includegraphics[width=\columnwidth]{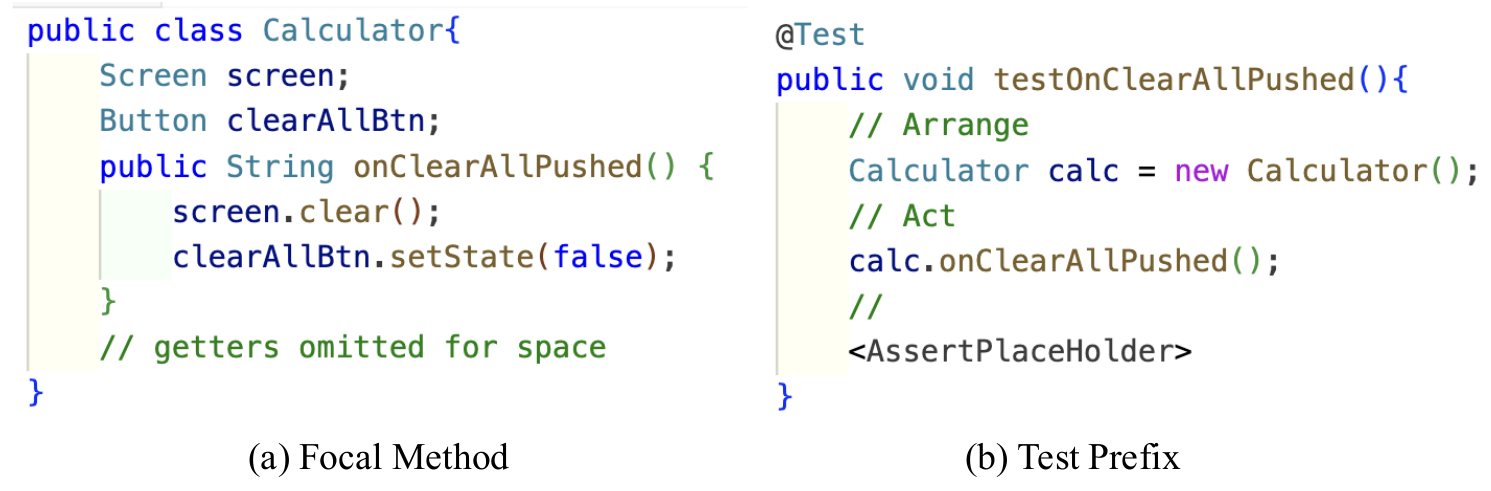}
    \caption{An illustrative example of focal method and test prefix}
    \label{fig:motivating-example}
\end{figure}
Generally, assertion generation techniques take the focal method (\textbf{$FM$}) and a test prefix (\textbf{$TP$}) as input, and generate one or more assertions for the \textbf{$FM-TP$} pair~\cite{IRAG}. 
Figure~\ref{fig:motivating-example}(a) and (b) illustrate the focal method \texttt{onClearAllPushed} and its corresponding test prefix \texttt{testOnClearAllPushed}, respectively.
The \texttt{onClearAllPushed} method defines the behavior when the clear-all button is pressed: it clears all the content in the \texttt{screen} field and updates the \texttt{state} of \texttt{clearAllBtn} to indicate activation (i.e., \texttt{false}). 
In the test prefix, the \texttt{testOnClearAllPushed} method sets up the testing scenario by instantiating a \texttt{Calculator} object and invoking the focal method.

Since the focal method performs two actions (clearing the screen and activating the button), comprehensive testing requires two corresponding assertions: one to verify the updated \texttt{state} and another to confirm the cleared \texttt{screen}.
Moreover, the form of assertions may vary depending on the testing approach.
For instance, to verify that the screen has been cleared, one may use \texttt{assertEquals} to compare the screen content with a literal string value ``0'', or alternatively, use \texttt{assertTrue} to validate the equivalence relationship, i.e., \texttt{content == "0"}.
Existing work often treats assertion generation as a unified task, which leads to an excessively large search space and insufficient guidance during exploration, particularly for advanced LLM-based approaches.
As a consequence, these methods become highly susceptible to the inherent randomness of LLMs, ultimately undermining their effectiveness. 
For example, the state-of-the-art method ChatAssert prompts ChatGPT to generate a single assertion per input. 
In doing so, the model must implicitly decide both what to test and what to expect in one step.
Given the wide range of possible test targets and the diverse ways to validate them, generating a correct assertion on the first attempt becomes extremely difficult, especially when limited to one-shot inference.
The top-1 response generated by ChatAssert for Figure~\ref{fig:motivating-example} is \texttt{assertFalse(calc.getClearAllBtn().getState\\());} which verifies the activation state of the clear-all button but fails to test whether the \texttt{screen} content has been cleared, leaving a key aspect of the behavior unverified.

To mitigate the challenge of undirected exploration in such a vast space, we propose a formal decomposition of the assertion into three distinct components: \textit{Assertion Function}, \textit{Expected Value}, and \textit{Actual Value}.
This fine-grained task partitioning provides a more structured framework for guiding the assertion generation process, enabling more targeted reasoning and reducing the impact of stochastic variability.
Figure~\ref{fig:terminology-assertion} presents an example assertion formulation using the JUnit framework. 
Related assertion terminologies are defined below:

\begin{figure}[ht]
    \centering
    \includegraphics[width=\columnwidth]{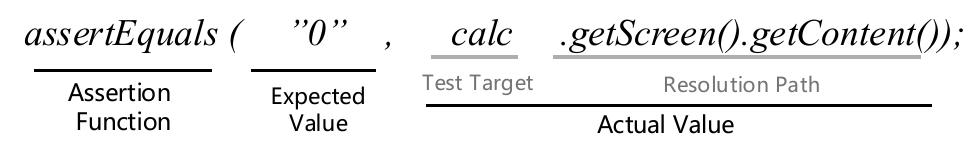}
    \caption{A fine-grained JUnit assertion for the illustration}
    \label{fig:terminology-assertion}
\end{figure}

\textbf{\textit{Assertion Function}} is the assertion method used in the test case, which defines the expected relationship between the actual value and the expected value. 
For instance, \texttt{assertEquals} is suitable for verifying whether the \texttt{screen} has been correctly set to ``0''.

\textbf{\textit{Actual Value}} is the target element being evaluated in the assertion. 
It can take various forms, including variables, method invocations, or combinations of both.
To more precisely describe its composition, we define the actual value as consisting of two components: the \textit{Test Target} and the \textit{Resolution Path}.
The test target refers to a variable (e.g., \texttt{calc}) defined within the test prefix, or a field defined in the test class. 
The resolution path represents a method invocation or a sequence of invocations applied to the test target, such as the \texttt{.getScreen().getContent()} method in Figure~\ref{fig:terminology-assertion}, which specifies the extent to which the test target is inspected.

\textbf{\textit{Expected Value}} represents the desired outcome of the actual value in an assertion. 
Unlike the actual value, which may have a complex structure, we define expected values in an atomic form.
Importantly, the expected value is not always a concrete data value; it can also involve selecting the appropriate assertion function that best expresses the intended condition, such as \texttt{assertTrue} for the equivalence relationship. 

\subsection{Test Assertion Generation Techniques}
As test oracles remain one of the most critical challenges in software testing, assertion generation has been widely studied in recent years~\cite{ATLAS,TOGLL,TOGA,IRAG,EditAS,ChatAssert}. Existing techniques can be categorized into three groups: traditional, deep learning-based, and large language model-based approaches.
Traditional techniques are typically divided into two major types: Symbolic Execution~\cite{DBLP:conf/relsoft/BoyerEL75,DBLP:journals/tse/Howden77,DBLP:conf/ibm/King74,DBLP:conf/tacas/KhurshidPV03,DBLP:conf/osdi/CadarDE08,DBLP:conf/kbse/PasareanuR10} and Specification Mining~\cite{DBLP:conf/kbse/ZhengMLXK11,DBLP:conf/icst/TanMTL12,DBLP:conf/issta/GoffiGEP16}.
The former explores program paths to derive logical conditions for assertions, while the latter extracts expected behaviors from documentation to guide assertion construction.
To address coverage, scalability, and adaptability challenges faced by traditional techniques, deep learning-based and language model-based techniques have gained increasing attention.

\smallskip
\noindent
\textbf{Deep Learning-Based Techniques.}
ATLAS~\cite{ATLAS} is one of the earliest studies to apply deep learning to assertion generation. It takes the focal method and test prefix as input and uses a recurrent neural network to generate assertions.
TOGA~\cite{TOGA} formulates assertion generation as a multi-class classification problem. It first constructs a dictionary of common values and types, then generates candidate assertions, and finally uses an Assertion Oracle Ranker to select the most suitable one.
Integrating information retrieval (IR) algorithm with deep learning models~\cite{IRAG,EditAS} takes the assertions as documents, and the inputs (i.e., focal method and test prefix) as queries, and uses lexical similarity (e.g., Jaccard Similarity) as the metric for searching.
Yu et al.~\cite{IRAG} were among the first to propose combining IR with deep learning for this task. They used IR to retrieve assertions similar to those required for the input and then modified the retrieved assertions to better match the target context.
EditAS~\cite{EditAS} further advanced this idea by explicitly modeling the differences between the retrieved and actual input pairs. It learns both edit operation sequences and token-level transformations to enhance the quality and accuracy of generated assertions.

DL models have significantly advanced assertion generation by leveraging their capabilities in code understanding and generation. However, these techniques require large amounts of labeled data for training and are prone to overfitting. 
Additionally, IR-based approaches are constrained by the similarity between retrieved and target assertions, limiting their adaptability to diverse code contexts.

\smallskip
\noindent
\textbf{Language Model-Based Techniques.}
Language models, trained on large-scale code corpora, have demonstrated strong generalizability across a wide range of code-related tasks~~\cite{DBLP:conf/issta/TianSWCK024, ChatTester,yang2025clarifying,UTLLMStudy}. 
TOGLL~\cite{TOGLL} extends TOGA by addressing generalizability challenges through fine-tuning a larger code model and demonstrates strong performance, particularly in mutation killing.
~\cite{zhang2025improving} proposed RetriGen, a hybrid approach that combines assertion retrieval with a pre-trained language model (PLM)-based assertion generator to improve assertion quality.
While LLMs have proven effective in generating human-readable unit tests~\cite{ChatTester}, their potential in assertion generation remains relatively underexplored.
Hayet et al.~\cite{ChatAssert} introduced ChatAssert, which guides LLMs using code summaries and one-shot examples to generate test assertions, followed by an ``execute-and-fix'' mechanism to correct invalid outputs. 

Although ChatAssert demonstrates the potential of LLM-based approaches over prior work, it relies on a single prompting strategy and is heavily dependent on runtime execution for post-hoc correction, which limits its reliability and scalability in practical settings.
In contrast, this study proposes an agent-based framework that integrates multiple prompting strategies in a structured and verifiable manner.
By decoupling key tasks (such as actual value construction and assertion formulation), and avoiding reliance on execution-based repair, our approach aims to improve both the accuracy and robustness of assertion generation in unit testing scenarios.
\subsection{LLM-Based Unit Test Generation}
Recent advances in LLMs have significantly accelerated end-to-end unit test generation, where LLMs generate complete test cases including test setup, method invocations, and assertions.
Representative approaches such as ChatTester~\cite{ChatTester} and ChatUniTest~\cite{ChatUnitest} have demonstrated impressive capabilities in automatically constructing executable unit tests with high coverage. These studies highlight the potential of LLMs to automate broader software testing workflows by jointly synthesizing test inputs and test oracles.
Nevertheless, complete unit test generation and test assertion generation address different challenges. End-to-end unit test generation primarily focuses on constructing executable tests that effectively exercise program behaviors, whereas assertion generation focuses on constructing precise test oracles that correctly validate those behaviors.
High code coverage alone does not guarantee effective fault detection if the generated assertions are weak, redundant, or semantically incorrect. Consequently, the quality of assertions often becomes the bottleneck in the effectiveness of automatically generated tests.
Therefore, we view these two research directions as complementary rather than competing. Assertion generation techniques can naturally serve as an oracle enhancement module for tests produced by traditional automated testing tools (e.g., EvoSuite and Randoop) as well as modern LLM-based unit test generation frameworks. Following this complementary view, AssertMate focuses specifically on oracle construction, with the goal of improving the correctness and fault-detection capability of automatically generated unit tests regardless of how the test prefixes are obtained.

\section{Methodology}
\label{methodology}
\begin{figure}
    \centering
    \includegraphics[width=\linewidth]{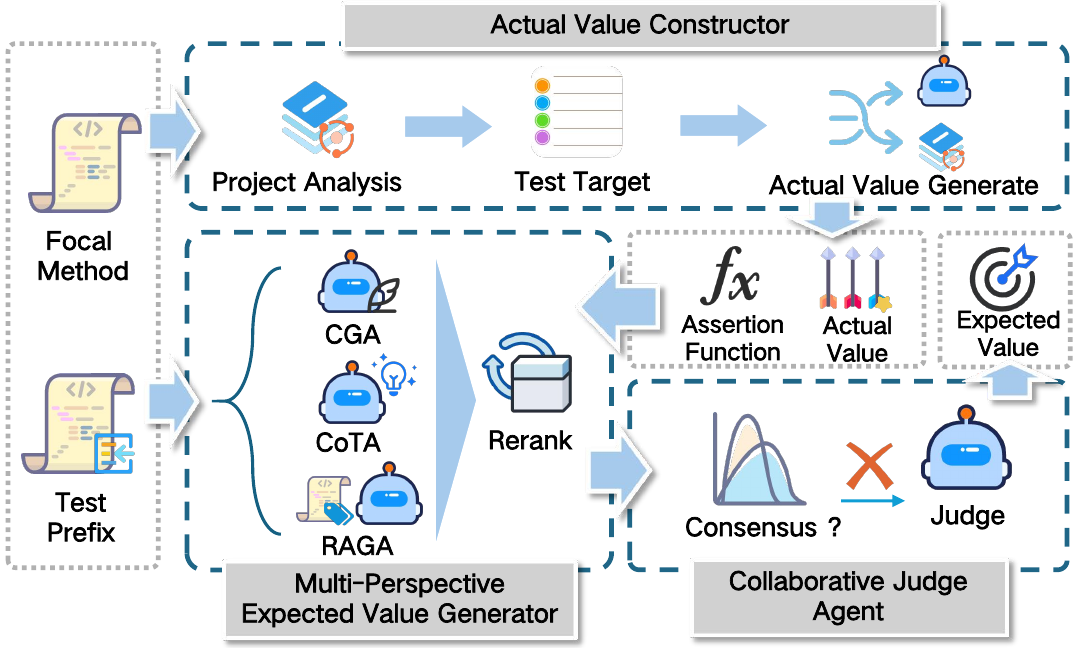}
    \caption{Overview of the agent-based test assertion generation framework \TECH{}}
    \label{fig:overview}
    \vspace{-15pt}
\end{figure}

Figure~\ref{fig:overview} presents an overview of \textbf{\TECH{}}. 
As shown, \TECH{} is composed of two main modules: the \textbf{Actual Value Constructor (\AVCAgent{}}) and the \textbf{Multi-Perspective Expected Value Generator (\MAGE{}}).
Given a focal method and its test prefix, \AVCAgent{} first identifies a set of actual values and corresponding assertion functions, using a hybrid rule-based approach that combines static program analysis with LLM assistance.
Subsequently, \MAGE{} first generates the matching expected values for each actual value identified by \AVCAgent{} though multi-perspective inference. 
Then, \MAGE{} elects the best-fitting output assertion through a multi-agent collaboration mechanism.
Each assertion is formed by combining the assertion function, actual value, and predicted expected value. Each generated assertion can be independently appended to the test prefix to produce a complete test case.
To make the workflow of \TECH{} more intuitive, we use the illustrative example shown in Figure~\ref{fig:motivating-example} as a running example throughout the remainder of this section.

\subsection{Actual Value Constructor}
To avoid overly-large search space when adopting multi-perspective inference, and also to mitigate the impact brought by randomness, it is necessary to first specify concrete and valid actual values as the target of multi-perspective reasoning. Therefore, \AVCAgent{} is designed to identify a set of valid actual values as a pre-processing step before multi-perspective reasoning.
Specifically, \AVCAgent{} implements 
a hybrid rule-based solution that combines static program analysis with LLM assistant to systematically identify the actual values that should be tested, thereby addressing the fundamental challenge of determining ``what to test.''

\AVCAgent{} begins by selecting appropriate test targets for assertions through static program analysis. 
Specifically, it identifies two basic test targets as a starter for the iterative traversing approach: (1) \textit{Return value of the focal method}, which is verified to ensure the method produces the expected output when it returns an object or value; and (2) \textit{Attributes defined in the focal class that have public getter methods}, since these attributes reflect side effects of the method’s execution and are verified to confirm that the intended modifications have been correctly applied.


Next, \AVCAgent{} constructs actual values through a type-aware iterative traversal process, supplemented by LLM assistance to handle external types. 
Given a set of candidate actual values, \AVCAgent{} applies a rule-based iterative analysis to generate the final actual values. 
The specific rules are designed based on the types of candidate actual values, enabling accurate and context-sensitive construction of assertion expressions.
We define the following three rules:

\begin{itemize}[leftmargin=10pt]
\item \textbf{Primitive data types} refer to a set of Java’s basic built-in types (e.g., \texttt{boolean}) as well as commonly used wrapper classes such as \texttt{String} and \texttt{Integer}, since their values can be directly compared. For these types, \AVCAgent{} includes the actual value in the final set without further decomposition or traversal.
\item \textbf{In-Project types} denote types defined within the project, such as \texttt{Button} and \texttt{Screen} in the motivating example. For these types, \AVCAgent{} expands the candidate actual value set by including attributes that have parameterless getter methods. 
Specifically, a getter method invocation is attached to the end of the resolution path and forms a new candidate value.

\item \textbf{External types} refer to types that are not defined within the project, such as those from third-party libraries or the Java Collections Framework.
Due to the lack of sufficient code context (e.g., missing class definitions), \AVCAgent{} leverages an LLM to assist in generating final actual values for these types.
Specifically, \AVCAgent{} queries the LLM to determine how to properly test a given actual value, especially for the resolution path. 
\end{itemize}

Based on the types of final actual values, \AVCAgent{} further determines the assertion functions to be used.
In this study, we focus on five common types of test assertions: \texttt{assertEquals}, \texttt{assertTrue}, \texttt{assertFalse}, \texttt{assertNull}, and \texttt{assertNotNull}.
These five types account for up to 86\% of all assertions, according to the empirical study conducted by Yu et al.~\cite{IRAG}, based on data collected from open-source repositories.
Specifically, for boolean values, \AVCAgent{} selects \texttt{assertTrue} or \texttt{assertFalse} as the appropriate assertion functions. For other primitive data types, it uses \texttt{assertEquals}. 
If a test target may be \texttt{null}, for example, when the focal method may return \texttt{null} under certain conditions, \AVCAgent{} includes an additional \texttt{assertNull} or \texttt{assertNotNull} assertion accordingly.
For types classified as external types, \AVCAgent{} consults the LLM to determine the appropriate assertion function based on the type and context of the actual value.



\textbf{Illustrative Example.} Using the focal method and test prefix from Figure~\ref{fig:motivating-example} as an example, 
\AVCAgent{} first identifies two attributes (i.e., \texttt{screen} and \texttt{clearAllBtn}) defined in the \texttt{Calculator} class as valid targets, and adds \texttt{calculator.getScreen()} and \texttt{calculator.getClearAllBtn()} to the initial candidate actual value set.
Since types (\texttt{Screen} and \texttt{Button}) are defined within the same project, \AVCAgent{} classifies them as In-Project types, inspects their class definitions, and further identifies \texttt{getContent()} and \texttt{getState()} as valid getter methods with a type of String and boolean, respectively. 
These are both considered primitive or directly comparable types, so the iteration terminates.
Finally, \AVCAgent{} constructs two actual values: \texttt{calculator.getScreen().getContent()} and \texttt{calculator.getClearAllBtn().getState()}, with types \texttt{String} and \texttt{boolean}, respectively.
The output of \AVCAgent{} for this case is:

\begin{lstlisting}[language=Java, backgroundcolor=\color{backcolour}, basicstyle=\ttfamily\footnotesize, xleftmargin=0pt]
assertEquals(<exp>,calculator.getScreen().getContent());
assert<exp>(calculator.getClearAllBtn().getState());
\end{lstlisting}
where $<$exp$>$ is the placeholder for expected value.


\subsection{Multi-Perspective Expected Value Generator}
\TECH{} employs a Multi-Perspective Expected Value Generator (\MAGE{}) to address ``what to expect'' and generate the corresponding expected values for test assertions.
\MAGE{} first introduces multiple prompting strategies, each acting as a distinct agent, to improve the comprehensiveness of expected value prediction. 
These agents independently generate candidate expected values based on the given context. 
Finally, \MAGE{} adopts a multi-agent collaboration mechanism to collectively select the most appropriate expected value.

\label{sec:mage}


\textbf{Multi-Perspective Agent Design.} 
\TECH{} leverages a multi-agent design to provide diverse perspectives for expected value generation, thereby mitigating the overconfidence and bias associated with relying on a single prompt. To achieve this, we surveyed prompting strategies used in recent work on assertion and unit test generation~\cite{ChatAssert,UTLLMStudy,ChatTester,ChatUnitest}, and identified three major categories of prompt design. Based on this analysis, we implemented three specialized assertion generation agents: the Code Generation Agent, the Retrieval-Augmented Generation (RAG) Agent, and the Chain-of-Thought (CoT) Agent.

\begin{itemize}[leftmargin=10pt]
    \item \textbf{Code Generation Agent (CGA).} Given that assertion generation is a sub-task of code generation and LLMs have demonstrated strong capabilities in this domain, the Code Generation Agent (CGA) leverages this strength by prompting the model to complete the test case with an appropriate assertion for a given actual value. This strategy offers a valuable perspective grounded in code structure and semantic consistency.
    \item \textbf{RAG Agent (RAGA).} IR-based assertion generation has shown promising results~\cite{IRAG, EditAS}, and Retrieval-Augmented Generation (RAG) is a well-established in-context learning technique in the LLM literature. Building on the Jaccard similarity-based retrieval algorithm used by EditAS~\cite{EditAS}, we design a RAG Agent (RAGA) for assertion generation. Specifically, the RAGA first retrieves the test case with the highest Jaccard similarity to the input test prefix, and then uses the retrieved case as a one-shot example to guide the LLM in generating an assertion for the given actual value. 
    This approach enables RAGA to effectively incorporate external knowledge and provides a unique perspective by supplying contextual reference for assertion generation.
    
    \item \textbf{CoT Agent (CoTA)}. Instructing LLMs to reason through the functionality of a given focal method has proven effective in prior work~\cite{ChatAssert,ChatTester}. 
    Building on this idea, we design a CoT Agent (CoTA) that applies the Chain-of-Thought (CoT) prompting strategy to break down complex inference tasks and support assertion generation through step-by-step reasoning.
    Specifically, CoTA operates in three main stages: 1) focal method understanding, 2) test scenario understanding, and 3) expected value prediction. Finally, CoTA synthesizes a test assertion by querying the LLM to produce a statement consistent with the reasoning chain accumulated throughout the session. 
    CoTA offers a unique perspective by summarizing the intended behavior of the focal method and the test design in the test prefix, thereby potentially improving assertion quality through deeper contextual understanding.
\end{itemize}




    
\begin{figure}
    \centering
    \includegraphics[width=0.9\columnwidth]{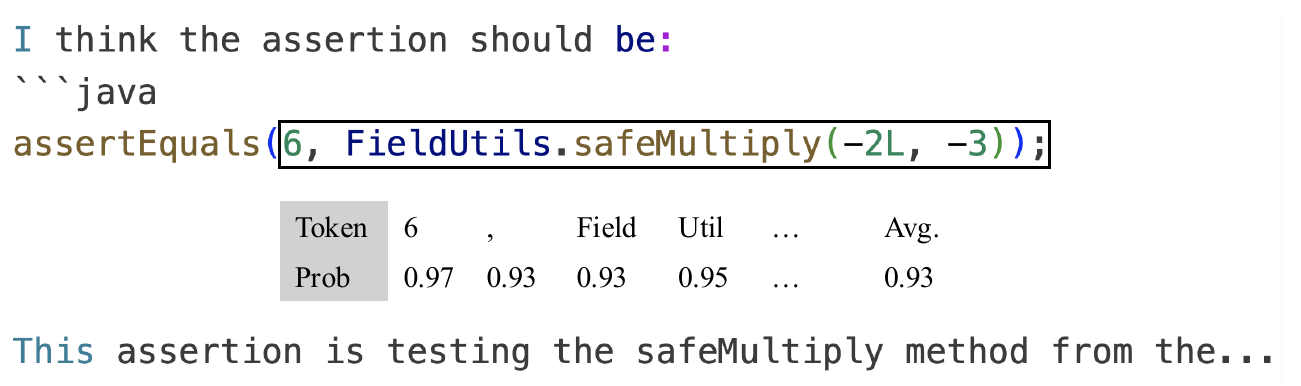}
    \caption{Example of LLMs' response tokens and probabilities}
    \label{fig:prob_example}
    \vspace{-10pt}
\end{figure}

\textbf{Prompt Prefill.}
In addition to designing effective agents, another major challenge lies in analyzing and constraining the responses of LLMs.
More recently, Yang et al.~\cite{UTLLMStudy} introduced a prefill mode for unit test generation. By prepending a predefined header to the LLM’s response, this mode reduces the likelihood of unexpected outputs and imposes structural constraints, such as restricting accessible classes, which helps improve the compilability of the generated code.
Inspired by this, \TECH{} also incorporates a prefill mechanism in each agent. 
Specifically, when \AVCAgent{} determines a specific assertion function (e.g., \texttt{assertEquals}), \TECH{} pre-attaches the corresponding function signature (e.g., \texttt{assertEquals}) at the beginning of the LLM's response to guide generation. In cases where \AVCAgent{} identifies multiple candidate assertion functions (e.g., \texttt{assertTrue} or \texttt{assertFalse}), \TECH{} attaches a more generic prefix such as \texttt{assert} to leave room for model inference while maintaining structural control.
For example, Figure~\ref{fig:prob_example} presents an illustrative case of an LLM-generated response. In this example, the prefilled content
``I think the assertion should be:\textbackslash n \texttt{```java}\textbackslash n \texttt{assertEquals(}''
is prepended to the LLM's response, and each agent in \MAGE{} begins generating from the position immediately following the left parenthesis.

\textbf{Probability-Based Response Re-Rank}.
To enhance the comprehensiveness of assertion generation, \MAGE{} adopts an oversampling strategy. Following ChatAssert~\cite{ChatAssert}, each agent generates five responses per input (i.e., FM-TP pair and actual value), resulting in 15 total candidates (3 components × 5 responses). 
Our choice of five candidates aligns with prior work~\cite{ATLAS,ChatAssert, primbs2025assert5}, where performance gains plateau beyond this point. 
Empirical results show notable improvements from top-1 to top-5, with diminishing returns from top-5 to top-10, balancing cost-effectiveness and effectiveness. However, this strategy introduces the challenge of confidently selecting the best candidate from a larger pool.

To address this, \TECH{} devises a probability-based re-ranking algorithm to prioritize responses with higher model confidence for each agent.
In LLMs, each token in the response is sampled from the vocabulary, and its associated probability reflects the model’s confidence in appending that token to the input.
Since LLM-generated responses often contain both the expected output (i.e., the assertion) and an accompanying explanation, incorporating the probabilities of explanatory tokens could bias the confidence estimation. 
To mitigate this, \TECH{} evaluates only the probabilities of tokens within the assertion portion of the response.
Figure~\ref{fig:prob_example} illustrates an example of an LLM-generated response along with token probabilities.
As shown, the response contains two parts, the assertion itself (i.e., \texttt{assertEquals(6, FieldUtils.safeMultiply(-2L,-3));}) and a short explanation \textit{``The assertion is testing the method...''}. 
In this case, \TECH{} computes the average probability (i.e., 0.93 in the ``Avg.'') and uses this value as the confidence score for the response. For each agent, the response with the highest confidence score is selected and forwarded to the subsequent collaboration stage.

\subsection{Collaborative Judge Agent}
\label{sec:judge}
After obtaining the top-ranked response from each agent, the next task is to determine which response best aligns with the given test prefix.
To accomplish this, \TECH{} adopts a multi-agent collaboration strategy to select the final expected value. 
When the multi-perspective expected value generators do not reach a consensus on how the assertion should be written, \TECH{} delegates the decision to a designated agent, referred to as \textit{Judge}, which conducts an in-depth analysis of the candidate responses and selects the best-fitting one.
\textit{Judge} employs a CoT reasoning strategy, as recommended in~\cite{LLMAsJudge}, to enhance its decision-making capabilities. 
Additionally, it utilizes the prefill mode to ensure structured and consistent evaluation of agent responses.
As an illustrative example, \TECH{} finally generates two assertions for the motivating case in Figure~\ref{fig:motivating-example} by predicting expected values based on actual values provided by \AVCAgent{}:
\begin{lstlisting}[language=Java, backgroundcolor=\color{backcolour}, basicstyle=\ttfamily\footnotesize, xleftmargin=0pt, escapeinside=``]
assertEquals(`\color{blue}\uline{\textbf{"0"}}`,calculator.getScreen().getContent());
assert`\color{blue}\uline{\textbf{False}}`(calculator.getClearAllBtn().getState());
\end{lstlisting}



All prompt designs used in \TECH{} are publicly available as part of our replication package~\cite{homepage} to facilitate reproducibility and future research.


\section{Experimental Setup}
\label{setup}
To evaluate the effectiveness of \TECH{}, we formulate the following three research questions to guide our study:
\begin{itemize}
    \item \textbf{RQ1:} How does \TECH{} perform compared to existing approaches?
    \item \textbf{RQ2:} How does \TECH{} perform when incorporated with Evosuite? 
    \item \textbf{RQ3:} How much contribution do \AVCAgent{} and \EVGenerator{} owe to the effectiveness of \TECH{}?
\end{itemize}

\subsection{Dataset Preparation}
\label{sec:dataset}

In this study, we utilize the Defects4J benchmark (v2.1.0)~\cite{DBLP:conf/issta/JustJE14} for evaluation. 
Although other datasets are available for assertion generation~\cite{ATLAS,TECO,method2test,TOGA,TOGLL}, none of them, unlike Defects4J, support evaluation of bug detection capability, which is essential for assessing the practical effectiveness and quality of generated assertions.
Defects4J includes 834 real-world bugs and has been widely adopted in prior studies~\cite{UTLLMStudy,TOGLL,li2024hybrid,jiang2023variable}.  Each bug is accompanied by a buggy version, a fixed version, and a set of bug-triggering test cases, enabling comprehensive and reproducible evaluation. 
Additionally, Defects4J provides several off-the-shelf tools, such as EvoSuite and Randoop, for automated unit test generation, further supporting the experimental process.
Specifically, we construct the following two sets of inputs, i.e., pairs of focal methods (FM) and test prefixes (TP), based on the bugs in Defects4J to evaluate the performance of \TECH{}.

\noindent \uline{\textbf{Human-Written FM-TP pairs.}} To evaluate the correctness of generated assertions, we first utilized the human-written test cases provided in Defects4J. These test cases offer a reliable foundation for assessing both pass rates and bug detection capabilities.

We first select bugs whose triggering tests fail at assertion statements. This criterion is based on the failure location during test execution, rather than the underlying fault type; therefore, we do not filter bugs based on whether they involve exceptions or other specific fault categories. 
Exception-related bugs are retained if the corresponding behavior is verified through an assertion and the triggering test successfully reaches that assertion. 
We exclude only cases where execution terminates before reaching the target assertion (e.g., due to an uncaught exception), because a generated assertion cannot be evaluated without an executable assertion context. Following this criterion, we obtain 112 eligible bugs for constructing FM–TP pairs.
%

Next, we identify the focal methods associated with the selected bugs by combining patch-level and test-level evidence. For each bug, we extract the methods modified by the bug-fixing patch and the methods invoked by the triggering test. When these two sets overlap, the corresponding method is selected as the focal method for constructing an FM–TP pair. Cases without such a reliable correspondence are excluded because both AssertMate and existing assertion-generation baselines require a focal method as input. This matching strategy avoids unreliable focal-method inference from complex test cases while leveraging the explicit root-cause information provided by bug-fixing patches.
To ensure practical relevance, we statistically analyzed the complexity of the selected assertion-triggering bugs. 
The results show that, on average, the selected bugs span 35 lines of code, involve 68 method calls, and have an average nesting depth of 9.
In cases where a single test method contains multiple assertions, we extract all assertion statements via program analysis and construct a distinct test case for each one.
Each synthesized test case contains all preceding code statements (including prior assertions) up to the current assertion, ensuring that necessary variable states are preserved.
Therefore, our construction does not assume that the oracle corresponds only to the final assertion of a test case; assertions appearing at intermediate positions are also included as independent targets.
The final assertion in each synthesized test case is then determined as the target test assertion, forming an FM–TP pair for training and evaluation.
As a result, we obtained a total of 667 FM–TP pairs.

\noindent \uline{\textbf{Evosuite-Generated FM-TP pairs.}}
Following prior work~\cite{TOGA,TOGLL}, we integrated \TECH{} with EvoSuite to demonstrate its practical applicability in our evaluation.  
Specifically, we selected focal methods based on the bug-fixing patches for each bug in Defects4J and used EvoSuite to generate corresponding test cases. 
The test generation process followed EvoSuite’s default configuration and relied on its built-in coverage criteria as the primary guidance.
Following previous research~\cite{shamshiri2015automatically, TOGA}, we addressed EvoSuite randomness by allocating 10 minutes per target and using coverage-guided generation to maximize test case quality.
To construct valid test prefixes for focal methods, we removed all assertions from the generated test cases.
In total, we obtained 2,203 Evosuite-generated FM-TP pairs for evaluation.


\subsection{Baselines}
We selected four representative types of state-of-the-art assertion generation techniques for comparison, covering DL-based and language model-based approaches.

\textbf{\EditAS{}} is an information retrieval (IR)-based assertion generation technique. It retrieves a similar FM-TP pair and learns from the edit sequence between the retrieved pair and the input, modifying the retrieved assertion accordingly. \EditAS{} leverages an attention-based LSTM encoder-decoder architecture~\cite{lstm} to perform the assertion generation process.

\textbf{\TOGLL{}} is a pre-trained language (PLM)-based assertion generation technique. It fine-tunes the CodeGen-350M model~\cite{DBLP:conf/iclr/NijkampPHTWZSX23} on the SF110 dataset to perform assertion generation.

\textbf{\RetriGen{}} is an enhanced retrieval-augmented assertion generation technique that uses a hybrid assertion retriever, which considers both lexical and semantic similarity to identify the most relevant test-assert pair from external codebases. This hybrid retriever combines a token-based retriever with an embedding-based retriever. 
The retrieved assertion is then combined with the original focal test and used as input to fine-tune a PLM-based assertion generator, implemented on top of the CodeT5 model.

\textbf{\textit{ChatAssert}} is an LLM-based assertion generation technique. It first prompts ChatGPT to generate a detailed code summarization of the test prefix. In addition, it retrieves a one-shot example based on semantic similarity using UniXCoder~\cite{DBLP:conf/acl/GuoLDW0022} to enhance in-context learning. Finally, it asks ChatGPT to generate an assertion for the given test prefix and applies an “execute-and-fix” post-processing step to resolve potential issues (e.g., compilation errors).
In this study, we focus specifically on assertion generation in the context of unit testing, where generating assertions without assuming function correctness is essential. 
To ensure fairness, we disabled the repair module in the original baseline, as its reliance on ``bug-free functions'' fundamentally contradicts the assumptions of unit testing. The adapted version is referred to as \textbf{\ChatAssert{}}.

\subsection{Metrics}
\label{sec:metrics}
Previous research mainly relied on metrics such as Exact Match, BLEU, and CodeBLEU to evaluate the effectiveness of assertion generation~\cite{ATLAS,IRAG,EditAS,TOGA,TOGLL,ChatAssert}.
However, we argue that these metrics are insufficient for fully capturing the correctness and practical utility.
On the one hand, Exact Match fails to account for semantically equivalent assertions that differ syntactically from the ground truth but exhibit identical behavior. 
On the other hand, minor textual differences may have a limited impact on BLEU or CodeBLEU scores, yet could lead to serious issues such as invalid syntax or compilation failures.
To address these limitations, our evaluation primarily adopts six comprehensive metrics, \textit{Compilation Success Rate, Pass Rate, Bug Detection Rate, Number of Killed Mutants}, \textit{Token Overhead}, and \textit{Average Time}.

Our evaluation incorporates two complementary metrics to assess assertion quality from both syntactic and semantic perspectives. First, since producing compilable code constitutes the fundamental capability of any code generation system, we propose the \textit{Compilation Success Rate (CSR)} metric to quantify syntactic correctness. 
This aligns with prior studies that employed the Exact Match metric as a proxy for syntax-level validity, which inherently depends on correct syntax.
For semantic validation, we introduce the \textit{Pass Rate (PR)} metric to evaluate functional correctness while mitigating false negatives arising from semantically equivalent assertions. 
To avoid conflating assertion quality with assertion quantity, we evaluate CSR and PR at the assertion-candidate level rather than at the test-case level.
For each FM-TP pair, if \TECH{} constructs $N$ actual values and therefore generates $N$ assertions, we give each baseline the same assertion budget by duplicating the same FM-TP input $N$ times and asking the baseline to generate $N$ assertions.
Each generated assertion is then independently appended to the corresponding test prefix and wrapped as a separate single-assertion test case.
In other words, an FM-TP pair with an assertion budget of $N$ produces $N$ independent test cases for each technique, each containing the same test prefix and one generated assertion.
The metrics are formally defined as: \textit{CSR} $= \frac{N_{compiled}}{N_{total}}$, \textit{PR} $= \frac{N_{passed}}{N_{total}}$, where $N_{compiled}$ denotes the number of test cases that compile successfully, $N_{passed}$ represents the number of test cases that execute successfully, and $N_{total}$ is the total number of independently constructed single-assertion test cases under the normalized assertion budget.

We employed a suite of metrics to systematically evaluate the bug-revealing capability of the generated assertions. The primary metric, \textit{Bug Detection Rate (BDR)}, quantifies the effectiveness in uncovering real-world bugs from the Defects4J benchmark. 
It is defined as the ratio of detected bugs ($B_{\text{detected}}$) to the total number of applicable bugs ($B_{\text{total}}$), using the combination of generated assertions and bug-triggering test prefixes.
To assess capability against synthetic defects, we adopted mutation testing metrics following established practice~\cite{TOGLL}: $\mathit{\#Covered\ Mutants}$ counts mutants executed by synthesized test cases, $\mathit{\#Killed\ Mutants}$ tallies mutants detected through assertion failures, and $\mathit{Kill\ Rate}$ ($\frac{\mathit{Killed}}{\mathit{Covered}}$) measures the detection efficiency. 
This triangulated evaluation strategy provides a comprehensive understanding of assertion quality across both real-world and artificially injected bugs.

When comparing LLM-based approaches, we measured computational efficiency through two key dimensions as proposed in~\cite{tian2025fixing}. 
\textit{Token Overhead} reflects resource consumption via the average token count (including both prompt and response) per query, while \textit{Average Time} captures temporal cost as the mean processing duration per FM-TP pair. 
These metrics collectively assess practical deployment feasibility in contemporary AI-powered testing pipelines.

\subsection{Implementation and Environment}

\textit{Studied LLM.} For \TECH{}, we used \textbf{DeepSeek-Coder-7B-Instruct (DC-7B)}~\cite{DC7B} as the backbone model. This choice was motivated by two main considerations: (1) prior research\cite{UTLLMStudy} has shown that DC-7B offers a favorable trade-off between cost and performance in unit test generation; and (2) we aim to demonstrate that \TECH{} can more effectively leverage the capabilities of LLMs for assertion generation, even when using a smaller-scale model.
Regarding the LLM inference configuration, we set the temperature to 1.0 to encourage response diversity, which aligns with our oversampling strategy for generating complementary candidate assertions. We set top-p to 1.0 to avoid additional probability truncation and set the maximum token limit to 4,000 to accommodate sufficiently long generations.

\smallskip
\noindent
\textit{Baseline Replication.} For the baseline techniques, we either replicated them using their publicly available implementations or re-implemented them strictly following the descriptions provided in their papers.
Specifically, for \TOGLL{}, we followed the original design using CodeGen-350M~\cite{DBLP:conf/iclr/NijkampPHTWZSX23} with the P5 input format, i.e., (\(TP+[sep]+FM\)) for \TOGLL{}. 
To ensure a fair comparison, we used the original training and testing datasets provided by the authors of both \EditAS{} and \TOGLL{}. We first reproduced their reported results to validate our setup and then retained their original hyperparameter configurations for evaluation.
For \RetriGen{}, we used the publicly available fine-tuned model parameters and replicated the experimental setup from the original paper. 
In this setup, the retrieved assertion is selected from the training set and combined with the focal method and test prefix to form the augmented input for generation~\cite{zhang2025improving}.
For \ChatAssert{}, we used the official implementation released by the authors~\cite{chatassertimpl}, but excluded the post-hoc repair component to align with the focus of this study.

\smallskip
\noindent
\textit{Environment.}
We implemented our pipeline using tree-sitter~\cite{tree-sitter} for static program analysis, using PyTorch 2.1.0~\cite{Pytorch} and Transformers 4.44~\cite{hf-transformer} to build the LLM runtime environment. Furthermore, we used the VLLM library~\cite{vllm} to accelerate model inference. All experiments were conducted on Ubuntu 20.04 LTS, equipped with an Intel Xeon Gold 6248R CPU, 512 GB RAM, and four NVIDIA A100 GPUs.

\section{Results}
\label{results}
In this section, we present our three research questions along with their corresponding analysis processes and results.

\subsection{RQ1: Effectiveness comparison among studied techniques}
\label{sec:rq1}



\begin{table}
\centering
\caption{Performance of Studied Assertion Generation Techniques on Human-Written test prefixes (RQ1)
}
\label{tab:rq1}
\resizebox{\linewidth}{!}{%
\begin{tabular}{lrrrrrrrrrr}
\toprule
\multirow{2}{*}{\textbf{Technique}} &
\multicolumn{2}{c}{\textbf{CSR (\%)}} &
\multicolumn{2}{c}{\textbf{PR (\%)}} &
\multicolumn{2}{c}{\textbf{BDR (\%)}} &
\multicolumn{2}{c}{\textbf{Token Overhead}} &
\multicolumn{2}{c}{\textbf{Avg. Time (s)}} \\
& \textbf{Mean} & \textbf{SD}
& \textbf{Mean} & \textbf{SD}
& \textbf{Mean} & \textbf{SD}
& \textbf{Mean} & \textbf{SD}
& \textbf{Mean} & \textbf{SD} \\
\midrule

\EditAS{}     & 18.06 & 0.24 & 17.71 & 0.27 & 0.00  & 0.00 & --          & --     & 0.37  & 0.01 \\
\TOGLL{}      & 12.05 & 0.66 & 11.83 & 0.73 & 0.00  & 0.00 & --          & --     & 0.81  & 0.03 \\
\RetriGen{}   & 21.59 & 0.23 & 6.43  & 0.25 & 0.00  & 0.00 & --          & --     & 0.23  & 0.01 \\
\ChatAssert{} & 50.21 & 1.83 & 39.81 & 2.04 & 37.32 & 1.82 & 1{,}609.67 & 103.72 & 13.27 & 0.43 \\
\TECH{}       & \textbf{76.78} & 0.90 & \textbf{60.09} & 1.07 & \textbf{46.25} & 1.75 & 5{,}284.70 & 117.36 & 15.31 & 0.83 \\
\bottomrule
\end{tabular}%
}
\end{table}

\Design{} To answer RQ1, which compares the effectiveness of \TECH{} with baseline techniques, we used the \textbf{Human-Written FM-TP pairs} (described in Section~\ref{sec:dataset}) as inputs and applied each studied technique to generate corresponding assertions. The generated assertions were then appended to the test prefixes to form complete test cases.
We evaluated these test cases using the Defects4J benchmark and reported three key effectiveness metrics: \textit{Compilation Success Rate (CSR)}, \textit{Pass Rate (PR)}, and \textit{Bug Detection Rate (BDR)}. Additionally, to assess computational efficiency, we measured and reported the \textit{Token Overhead} and \textit{Average Time} for each technique.
To mitigate stochasticity in LLM operations and improve result reliability, we independently ran \TECH{} and every baseline five times with the same configuration and reported the per-technique mean and sample standard deviation (SD) for each metric. 
To assess the robustness of the observed improvements, we used Wilcoxon signed-rank tests~\cite{Woolson2007Wilcoxon} on paired outcomes after averaging the five repetitions. We additionally reported Cliff's delta ($\delta$)~\cite{Cliff1993Dominance} as a measure of effect size.
The evaluation results are summarized in Table~\ref{tab:rq1}.

\ResultAndAnalysis{}
Across five independent runs, \TECH{} achieves mean CSR, PR, and BDR values of 76.78\%, 60.09\%, and 46.25\%, respectively, with corresponding SDs of 0.90, 1.07, and 1.75 percentage points, demonstrating stable performance across runs. 
Overall, \TECH{} consistently outperforms the studied deep-learning-based and LLM-based approaches.
First, in terms of syntactic correctness, \TECH{} achieves a 52.92\% relative improvement in CSR over the best-performing baseline, \ChatAssert{} (50.21\%).
This substantial gain highlights the effectiveness of combining static program analysis with LLMs via \AVCAgent{}.
Second, for semantic correctness, \TECH{} delivers a 50.94\% relative gain in PR over \ChatAssert{} (39.81\%) likely due to the multi-agent collaboration strategy that integrates diverse prompting styles for improved semantic reasoning.
Compared with the other approaches, \TECH{} achieves 3.39$\times$, 5.08$\times$, and 9.35$\times$ higher PR of \EditAS{} (17.71\%), \TOGLL{} (11.83\%), and \RetriGen{} (6.43\%), respectively.
Third, in bug detection, \TECH{} attains the highest BDR on the Defects4J benchmark, compared with 37.32\% for \ChatAssert{}, while \EditAS{}, \TOGLL{}, and \RetriGen{} failed to detect any bugs, yielding a BDR of 0\%.

\begin{table}[t]
\centering
\caption{Performance of studied assertion generation techniques on Evosuite-generated test prefixes (RQ2)}
\label{tab:rq2}

\scriptsize
\begin{threeparttable}
\begin{tabular*}{\columnwidth}{@{\extracolsep{\fill}}lrrrr}

\toprule
\textbf{Technique} &
  \textbf{CSR (\%)} &
  \begin{tabular}[c]{@{}c@{}}\textbf{\# Covered}\\ \textbf{Mutants}\end{tabular} &
  \begin{tabular}[c]{@{}c@{}}\textbf{\# Killed}\\ \textbf{Mutants}\end{tabular} &
  \begin{tabular}[c]{@{}c@{}}\textbf{Kill}\\ \textbf{Rate (\%)}\end{tabular} \\
\midrule
\EditAS{}                 & 65.15  & 17{,}883 & 2{,}071 & 11.58 \\
\TOGLL{}                  & 70.95  & 29{,}224 & 5{,}374 & 18.39 \\
\textit{\TOGLL{}$_{DN}$}\tnote{*} & 1.21 & 138 & 9 & 6.52 \\
\RetriGen{}               & 35.54  & 973      & 196     & \textbf{20.14} \\
\ChatAssert{}             & 79.08  & 63{,}665 & 8{,}738 & 13.72 \\
\textit{EvoSuite}         & \textbf{100.00} & 56{,}661 & 8{,}184 & 14.44 \\
\TECH{}                   & 83.72  & \textbf{76{,}840} & \textbf{12{,}276} & 15.98 \\

\bottomrule
\end{tabular*}

\begin{tablenotes}
\item[*] \textit{\TOGLL{}$_{DN}$} refers to the variant of \TOGLL{} retrained on the unseen dataset.
\end{tablenotes}
\end{threeparttable}
\end{table}

The statistical testing results further support the effectiveness of \TECH{} over all baselines. Using the Wilcoxon signed-rank test with Cliff’s delta effect size analysis, we observe statistically significant improvements for all pairwise comparisons in terms of CSR, PR, and BDR.
For the comparison between \TECH{} and the best-performing \ChatAssert{}, \TECH{} achieves a significant improvement in CSR ($p=2.974\times10^{-29}$, $\delta=0.377$, medium effect), PR ($p=9.383\times10^{-22}$, $\delta=0.302$, small effect) and BDR ($p=0.0489$, $\delta=0.152$, small effect). These results demonstrate that the performance gains of \TECH{} are statistically significant across the evaluated baselines.

Efficiency analysis of computational costs reveals nuanced insights into \TECH{}'s operational trade-offs. While our approach requires 3.28× more tokens per generation than \ChatAssert{} (5,284.70 vs. 1,609.67 tokens), primarily due to multi-prompt integration and quality-oriented oversampling, this is offset by three key advantages: (1) The open-source DC-7B model outperforms ChatGPT in CSR (76.78\% vs. 50.21\%), showing that targeted architecture can enable small-scale models to exceed larger ones;

(2) Running 7B models is more cost-effective and practical than proprietary models like ChatGPT;
(3) Multi-agent parallelism keeps generation time comparable to ChatAssert (15.31s vs. 13.27s per generation), especially considering AssertMate's superior effectiveness.
This balanced cost-performance profile makes \TECH{} a compelling option for assertion generation in safety-critical or resource-constrained environments.
\begin{tcolorbox}[colback=gray!5,colframe=awesome]
    \textbf{RQ1 Summary}: 
    \TECH{} demonstrates strong assertion generation performance, achieving mean CSR, PR, and BDR values of 76.78\%, 60.09\%, and 46.25\%, respectively. The improvements over the baselines are statistically significant, corresponding to relative gains of \textbf{52.92\%--537.18\%} in CSR and \textbf{50.94\%--834.53\%} in PR.
    While consuming 3.28$\times$ more tokens than the state-of-the-art LLM-based approach \ChatAssert{}, \TECH{} maintains a practical trade-off with a controlled time and efficient performance using a small-scale open-source model.
\end{tcolorbox}

\subsection{RQ2: Integration with Evosuite}

\Design{} Following prior work~\cite{TOGLL}, we integrated the studied assertion generation techniques with EvoSuite to evaluate their practical utility. 
Specifically, we used the \textbf{EvoSuite-generated FM-TP pairs} (described in Section~\ref{sec:dataset}) as inputs and followed the same procedure as in RQ1 to construct complete test suites. We then employed the mutation testing toolkit provided by Defects4J to assess the effectiveness of each technique in killing mutants. Our evaluation focused on Compilation Success Rate (CSR), the number of covered and killed mutants, and the kill rate. We did not report Pass Rate or Bug Detection Rate due to the lack of valid ground-truth expected outputs in the generated test cases, which makes distinguishing false positives infeasible in the unit testing scenario. 
The results are presented in Table~\ref{tab:rq2}.

\smallskip
\ResultAndAnalysis{} 
Evaluation results demonstrate that \TECH{} represents an advancement in assertion generation integrated with EvoSuite, substantially outperforming baselines across both syntactic and semantic dimensions.
In terms of syntactic validity, \TECH{} achieves an 83.72\% CSR, surpassing \EditAS{}, \TOGLL{}, \RetriGen{}, and \ChatAssert{} by 28.50\%, 18.00\%, 135.57\%, and 5.87\%, respectively.
Semantically, \TECH{} exhibits outstanding assertion quality, killing 12,276 mutants, a 492.76\%, 128.43\%, and 40.49\% improvement over \EditAS{}, \TOGLL{}, and \ChatAssert{}, respectively. 
While \RetriGen{} achieves the highest kill rate, it covers the fewest mutants and kills far fewer mutants than the other compared techniques.
These results together underscore \TECH{}'s ability to generate precise expected values, which are essential for detecting subtle faults in complex codebases.

When compared to the classic generation tool EvoSuite, its extremely high CSR (100\%) is expected, as EvoSuite’s evolutionary process inherently ensures that the generated tests compile successfully on the fixed program versions.
In contrast, language model-based generators, including AssertMate, naturally face risks of syntactic hallucination, which can potentially reduce CSR. Nonetheless, AssertMate achieves the highest CSR among the studied assertion-generation techniques,
effectively suppressing hallucinations relative to other baselines such as ChatAssert*. 
More importantly, AssertMate achieves substantially higher mutant coverage (76,840 vs. 56,661), kill counts (12,276 vs. 8,184), and kill rate (15.98\% vs. 14.44\%),
demonstrating stronger fault-detection capability than EvoSuite.

Despite these advantages, our evaluation also reveals areas for refinement. While \TECH{} leads in the absolute number of killed mutants, its kill rate (15.98\%) suggests room for improving the precision of generated assertions, especially in capturing subtle behavioral variations.
We first note that overall, language model-based assertion generation approaches outperform the DL-based one (i.e., 11.58\% achieved by \EditAS{}).
Among the language model-based baselines, \RetriGen{} achieved the highest kill rate at 20.14\%; however, it covered the fewest mutants, with only 973 detected.
On the other hand, \TOGLL{} shows a kill rate (18.39\%), which may also superficially suggest superior assertion quality compared with \TECH{}.
However, our follow-up investigation uncovered a potential leakage issue: \TOGLL{} was originally trained on the SF110 dataset~\cite{SF110}, which includes EvoSuite-generated test cases. 
This overlap with the evaluation setup likely inflated its performance.
To test this hypothesis, we retrained \TOGLL{} using $Data_{new}$ dataset collected by Yu et al~\cite{IRAG}
and observed a sharp decline in performance, with the kill rate dropping from 18.39\% to 6.52\% (as shown in Table~\ref{tab:rq2}, line \TOGLL{}$_{DN}$). 
This confirms the susceptibility of data-driven PLM-based methods to training–test distribution alignment and highlights a key limitation where real-world assertion generation tasks rarely have access to aligned test data during training.
Together, these findings indicate that existing approaches relying on fine-tuning exhibit limited generalizability.

\begin{table}[]
\centering
\tabcolsep=4pt
\renewcommand\arraystretch{1.25}
\caption{Performance of studied variants (RQ3)
}
\label{tab:rq3}
\resizebox{\columnwidth}{!}{%
\begin{tabular}{@{}c|ccc|cc|cc|c@{}}
\toprule
\multirow{2}{*}{\textbf{Metric}} &
  \multicolumn{3}{c|}{{\ul \textbf{Single Agent}}} &
  \multicolumn{2}{c|}{{\ul \textbf{Collaboration}}} &
  \multicolumn{2}{c|}{{\ul \textbf{\AVCAgent{}}}} &
  \multirow{2}{*}{\textbf{\TECH{}}} \\
    & CGA    & RAGA   & CoTA   & Debate & Voting & \TECH{}$^{-}$ & \ChatAssert{}$^{+}$ &       \\ \hline
CSR & 52.01 & 61.46 & 70.74 & 73.53 & 75.32 & 12.46 & 51.24 & \textbf{76.78} \\
PR  & 43.96 & 41.79 & 46.59 & 53.25 & 44.73 & 10.27 & 52.17 & \textbf{60.09} \\
BDR & 38.39 & 35.71 & 41.07 & 45.53 & 37.50 & 0.00 & 48.21 & \textbf{46.25} \\ \bottomrule
\end{tabular}%
}
\vspace{-15pt}
\end{table}


\begin{tcolorbox}[colback=gray!5,colframe=awesome]
    \textbf{RQ2 Summary}: \TECH{} demonstrates strong practicality and superiority when integrated with EvoSuite, achieving substantially higher Compilation Success Rate (CSR) as well as greater numbers of covered and killed mutants. For example, \TECH{} attains an assertion-level CSR of 83.72\%, outperforming the studied assertion-generation approaches by margins ranging from 5.87\% to 135.57\%.
    Meanwhile, it highlights the chance to improve assertion specificity and generalizability.
\end{tcolorbox}

\subsection{RQ3: Ablation Study}
\label{sec:ablation}
\Design{} In this RQ, we further conducted an ablation study to investigate the extent to which each major component contributes to the effectiveness of \TECH{}.
Three sets of variants are designed: 
\begin{itemize}
    \item[1] \textbf{\AVCAgent{} Variants:} We assessed the impact of the actual value constructor (\AVCAgent{}) through two complementary experiments. 
    First, we removed \AVCAgent{} from \TECH{}, denoted as \TECH{}$^{-}$, to evaluate its direct contribution to \TECH{}'s overall effectiveness. Second, we integrated \AVCAgent{} into \ChatAssert{}, referred to as \ChatAssert{}$^{+}$, to demonstrate the generalizability of \AVCAgent{}.
    
    \item[2] \textbf{\MAGE{} Variants:} One key strength of \TECH{} lies in its multi-perspective expected value generator (\MAGE{}), which integrates diverse prompting strategies for expected value generation. To evaluate the contribution of this design, we compared the effectiveness of each individual agent, Code Generation Agent (CGA), Retrieval-Augmented Generation Agent (RAGA), and Chain-of-Thought Agent (CoTA), against the full \TECH{} version.

     \item[3] \textbf{Collaboration Strategies:} In this set of variants, inspired by prior research~\cite{debate,agentsurvey}, we investigated the effectiveness of another two representative agent collaboration strategies: \textbf{Debate} and \textbf{Voting}. 
     Debate involves iterative rounds of interaction where agents revise their responses based on peer input until a consensus is reached. 
     Voting is one of the most commonly used and fundamental decision-making strategies in multi-agent discussion scenarios and in this study, we used it as the naive strategy for comparison.
\end{itemize}


We used \textbf{Human-Written FM-TP pairs} as inputs and evaluated each variant's performance using Compilation Success Rate (CSR), Pass Rate (PR), and Bug Detection Rate (BDR). The results are presented in Table~\ref{tab:rq3}.

\smallskip
\ResultAndAnalysis{} 
The dual capabilities of \AVCAgent{} (reasoning guidance and hallucination mitigation) are key to its effectiveness in promoting LLM-based assertion generation. 
By leveraging static program analysis to construct accurate actual values, \AVCAgent{} reduces hallucinated outputs, as evidenced by the 64.32-percentage-point drop in CSR and complete loss of bug detection when \AVCAgent{} is removed (\TECH{}$^{-}$) compared to the full \TECH{}. Additionally, \AVCAgent{} enhances reasoning quality by anchoring assertion generation to concrete program states, leading to a PR improvement from 10.27\% in \TECH{}$^{-}$ to 60.09\% in \TECH{}.
Its adaptability is further confirmed by integrating it into \ChatAssert{}: \ChatAssert{}$^{+}$ outperforms the original \ChatAssert{} with relative gains of 2.05\%, 31.05\%, and 29.18\% in CSR, PR, and BDR, respectively. These results position \AVCAgent{} as both an effective hallucination safeguard and a transferable reasoning scaffold for LLM-based assertion generation techniques.

The \MAGE{} framework outperforms individual prompting agents (CGA, RAGA, and CoTA), as demonstrated by \TECH{}'s consistent superiority across all evaluation metrics.
While each prompting strategy captures distinct facets of assertion generation, their isolated use limits holistic coverage. By enabling collaboration among specialized agents, \MAGE{} integrates complementary insights and facilitates mutual validation. This coordinated design leads to relative improvements of 8.54\% in CSR, 28.98\% in PR, and 12.61\% in BDR over the best-performing standalone prompting strategy, demonstrating that structured multi-agent collaboration more effectively aligns generated assertions with both syntactic correctness and semantic intent.


The ``LLM-as-a-Judge'' serves as an effective collaboration mechanism, striking a balance between efficiency and effectiveness.
In contrast, the Debate mechanism, despite yielding a moderate bug detection rate (45.53\% BDR), suffers from the cascading effect of hallucinations across multi-round interactions (73.53\% CSR)~\cite{agentsurvey}.
These extended dialogues also incur prohibitive computational costs, consuming 3.7× more tokens per assertion than \TECH{}.
While voting achieves reasonable syntactic correctness (75.32\% CSR), manual inspection revealed that it inevitably discards 12.5\% of bugs detectable only by individual agents using specific prompting strategies (4 by CG, 3 by RAG, and 7 by CoT).
This inherent limitation of Voting accounts for its 8.75-percentage-point drop in BDR compared to \TECH{}.
\TECH{} overcomes these limitations through single-round expert mediation, where a \textit{Judge} agent preserves valuable minority perspectives while mitigating hallucination risks. By combining consensus validation with expert appraisal, \TECH{} achieves superior bug detection (46.25\% BDR) without sacrificing implementation stability.

Beyond comparisons with individual agents, Debate, and Voting, we further evaluate the selection effectiveness of the \textit{Judge} through a paired override analysis against CoTA, the strongest individual agent in Table III. 
Specifically, we compare the outcomes of the assertions selected by the \textit{Judge} and CoTA for the same input instance. 
A \textit{correct override} occurs when the CoTA-generated assertion fails but the \textit{Judge}-selected assertion succeeds, whereas an \textit{incorrect override} occurs when the opposite happens. 
The \textit{Judge} achieves 103 correct overrides and 44 incorrect overrides for CSR, and 129 correct overrides and 40 incorrect overrides for PR; both differences are statistically significant according to McNemar's test ($p<0.001$)~\cite{mccrum2008correct}. 
For BDR, the \textit{Judge} achieves 13 correct overrides and 4 incorrect overrides, also showing a statistically significant difference ($p<0.05$). These results demonstrate that the \textit{Judge} more frequently improves upon the strongest individual agent than incorrectly replacing a better candidate.


\begin{tcolorbox}[colback=gray!5,colframe=awesome]
    \textbf{RQ3 Summary}: Both \AVCAgent{} and \MAGE{} are critical to \TECH{}'s effectiveness. 
    \AVCAgent{} not only mitigates hallucinations and guides LLM reasoning during assertion generation, but also enhances existing LLM-based techniques when integrated. Additionally, the ``LLM-as-a-Judge'' strategy proves to be a more effective collaboration mechanism compared to debate and voting, striking a balance between effectiveness and efficiency.
\end{tcolorbox}
\section{Discussion}
\label{sec:discussion}
In this section, we qualitatively demonstrate the effectiveness of \TECH{} through a representative case, examine its performance with more advanced LLMs, discuss the complementarity and boundaries of its multi-agent design, and highlight the threats to validity of our study.

\subsection{Case Study of Test Assertion Quality}

\begin{figure}[htbp]
    \centering

    \begin{subfigure}{\linewidth}
        \centering
        \includegraphics[width=\linewidth]{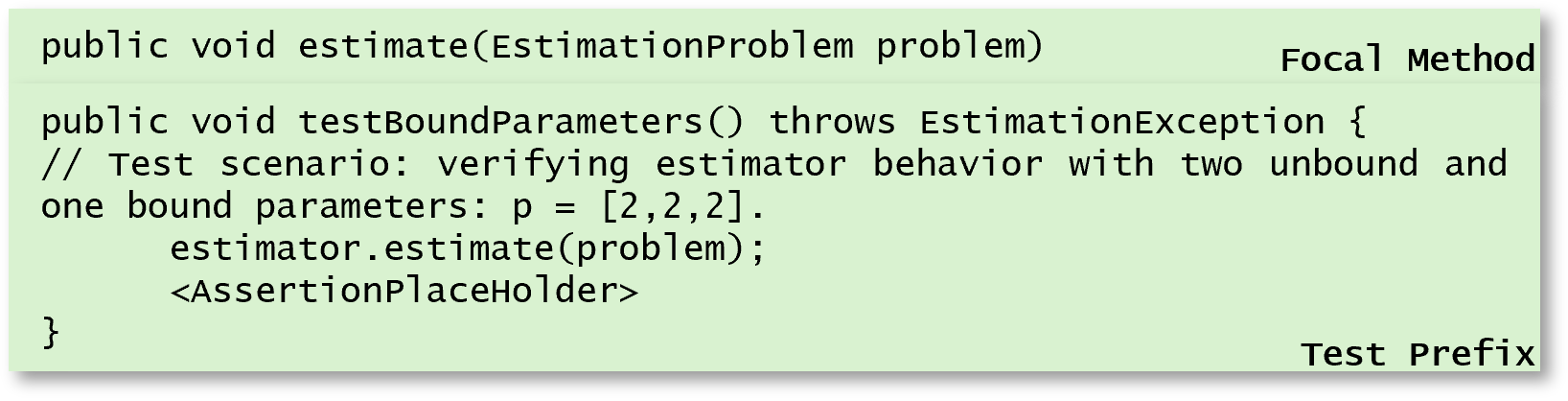}
        \caption{Focal Method and Test Prefix}
        \label{fig:cga_prompt}
    \end{subfigure}

    \begin{subfigure}{\linewidth}
        \centering
        \includegraphics[width=\linewidth]{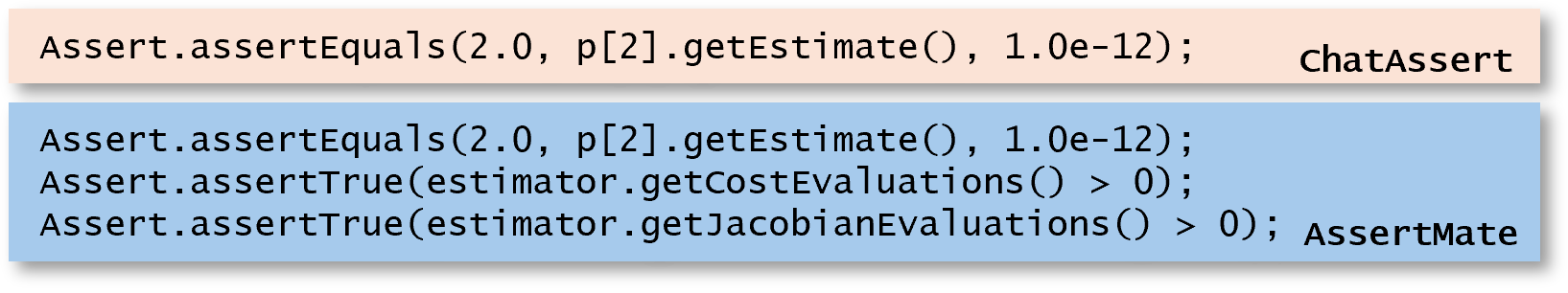}
        \caption{Test Assertions generated by \textit{AssertMate} and \ChatAssert{}}
        \label{fig:cota_prompt}
    \end{subfigure}

    \caption{A representative case study of bound parameter estimation}
    \label{fig:test_instance}
\end{figure}

Our evaluation (RQ1 and RQ2) confirms the effectiveness of the proposed \TECH{} in comparison to existing baselines, particularly the second-best performing LLM-based approach \ChatAssert{}. 
To gain deeper insights into their behavioral differences, we conducted an in-depth case study on a representative test instance from our dataset, as illustrated in Figure~\ref {fig:test_instance}, focusing on a scenario involving bound parameter estimation within an optimization problem.

\smallskip
\noindent
\textbf{Case Description.}
As shown in the figure, the method under test is a Gauss–Newton estimator for nonlinear least squares optimization, and the test constructs an estimation problem with two free parameters and one bound parameter.
The human-written ground truth is:
\begin{lstlisting}[language=Java, backgroundcolor=\color{backcolour}, basicstyle=\ttfamily\footnotesize, xleftmargin=0pt, escapeinside=`]
Assert.assertTrue(estimator.getRMS(problem) < 1.0e-10);
\end{lstlisting}
which verifies that the optimizer achieves a sufficiently small residual, indicating overall convergence of the estimation process.

\smallskip
\noindent
\textbf{Outputs of \ChatAssert{} and \TECH{}.}
To ensure an equal assertion budget, we invoked \ChatAssert{} three times for this case, matching the three assertions generated by \TECH{}.
However, all three generations yielded the same assertion shown in Figure~\ref{fig:test_instance}, which verifies that the bound parameter remains fixed at its initial value (2.0), capturing constraint enforcement but not overall convergence.
On the other hand, \TECH{} not only reproduced the parameter-invariance assertion but also generated additional semantically relevant ones.
This is because \AVCAgent{} detected additional internal attributes of the estimator class as potential actual values through static program analysis, enabling richer assertion generation.
Moreover, if we manually add public utility methods (such as getRMS()) of the target class into the input, our approach is also able to synthesize assertions like:
\begin{lstlisting}[language=Java, backgroundcolor=\color{backcolour}, basicstyle=\ttfamily\footnotesize, xleftmargin=0pt, escapeinside=`]
Assert.assertTrue(estimator.getRMS(problem) < 1.0e-3);
\end{lstlisting}
which checks the same convergence property as the human-written ground truth, although with a looser threshold.

This case study highlights the key advantages of \TECH{}:
By identifying a set of valid actual values for assertions, \TECH{} is able to target more meaningful verification points beyond the values of local variables, thereby enhancing the thoroughness of test validations.
At the same time, the example also uncovers opportunities for future improvement.
While \TECH{} leverages program structure to guide the selection of meaningful targets, it currently lacks the ability to automatically recognize and incorporate public utility methods (such as getRMS()), which often encapsulate essential correctness conditions.
This limitation suggests a promising research direction is optimizing the automated identification and inclusion of relevant public methods from the target class. 
Such methods usually serve as ideal actual values, and incorporating them could significantly improve the quality and relevance of generated assertions.

\subsection{\TECH{} with More Advanced LLMs}
Considering experimental costs and the practical usage, we initially adopted DC-7B as the base model for \TECH{}. 
However, with the rapid evolution of LLMs, more advanced models have demonstrated notably stronger capabilities in code understanding and generation. 

In particular, recent reasoning-enhanced models such as DeepSeek-R1 have garnered substantial attention for their superior performance in complex tasks.
To explore the potential benefits of integrating more advanced models into our framework, we conducted an additional experiment evaluating the effectiveness of \TECH{} when paired with larger and more advanced LLMs. 
Specifically, we selected two open-source models:  \textit{Qwen2.5-Coder-32B (QC-32B)}, representing state-of-the-art code-centric LLMs, and \textit{DeepSeek-R1-Distill-Qwen-32B (DR-32B)}, exemplifying reasoning-enhanced LLMs.
All experiments were conducted under the same setup as RQ1, and the corresponding results are presented in Table~\ref{tab:discussion}.

\textbf{Code-centric advanced LLMs tend to improve assertion generation effectiveness.}
Comparing the effectiveness of QC-32B and DC-7B, we observed that advanced code-centric LLMs significantly enhance assertion generation. 
Although QC-32B incurred a slight drop in CSR of 2.48 percentage points compared with DC-7B, it delivered gains of 8.64 and 11.79 percentage points in PR and BDR, respectively, highlighting its strength in generating semantically accurate and bug-revealing assertions.
These results suggest that, when resources permit, adopting larger and more advanced code LLMs can be a valuable choice for boosting assertion quality.


\begin{table}[]
\centering
\caption{Performance of using different base model for \TECH{}}
\label{tab:discussion}
\resizebox{0.7\columnwidth}{!}{%
\begin{tabular}{@{}lccc@{}}
\toprule
\textbf{Model}                      & \textbf{CSR (\%)}            & \textbf{PR (\%)}             & \textbf{BDR (\%)}            \\ \midrule
DR-32B    & 59.44          & 55.10          & 45.54          \\
QC-32B         & 74.30          & \textbf{68.73}  & \textbf{58.04} \\
DC-7B & \textbf{76.78} & 60.09          & 46.25       \\ \bottomrule

\end{tabular}%
}
\vspace{-15pt}
\end{table}

\textbf{While reasoning capabilities enhance LLM performance, they are not the sole determinant of assertion generation effectiveness.}
Although DR-32B offers advanced reasoning capabilities, it does not outperform DC-7B under the unified assertion-generation protocol adopted in this work, with CSR, PR, and BDR lower by 17.34, 4.99, and 0.71 percentage points, respectively.
One possible explanation is that the partially prefilled assertion template, while effective for ensuring executable and consistent outputs, may not be optimal for reasoning-oriented models, whose capabilities are better exploited through a more flexible think-then-answer generation process.
Nevertheless, DR-32B achieves the best bug detection performance on the Math project, suggesting that reasoning-oriented models remain advantageous for reasoning-intensive scenarios.
This observation motivates future work on adaptive prompting strategies that decouple semantic reasoning from structured assertion generation, allowing reasoning-oriented models to first reason about expected behaviors before producing executable assertions. More broadly, future agent collaboration strategies could dynamically route or weight different model families according to task characteristics, enabling reasoning-oriented and code-centric models to complement each other.


\subsection{Complementarity and Boundaries of Multi-Agent Collaboration}

The effectiveness of multi-perspective collaboration does not originate from the number of agents, but from the diversity of evidence sources and reasoning pathways they provide. AssertMate adopts CGA, RAGA, and CoTA because they capture complementary aspects of assertion reasoning. CGA derives expected values primarily from local program semantics and executable code structure; RAGA leverages structurally analogous tests to transfer assertion patterns and expected behaviors; and CoTA performs explicit multi-step reasoning by composing semantic relations from the focal method and test prefix. These heterogeneous perspectives reduce reliance on a single reasoning strategy and provide opportunities to identify valid assertions that may be missed by individual agents.

The RQ3 results provide empirical support for this complementarity. The agent ablation analysis demonstrates that different perspectives likely capture different aspects of assertion generation, while the collaboration-strategy comparison shows that effective aggregation can further improve the reliability of generated assertions. These results suggest that the benefit of multi-agent collaboration comes from combining complementary reasoning signals rather than simply increasing the number of agents.
However, multi-perspective collaboration is not universally beneficial. Its effectiveness depends on several conditions: (1) at least one perspective produces a correct candidate assertion, (2) different perspectives provide complementary rather than redundant evidence, and (3) the aggregation mechanism can identify and preserve valuable candidates. Collaboration may provide limited benefits when all agents lack necessary contextual information, share similar failure modes, or when the aggregation strategy eliminates useful minority candidates. 

The current evaluation explores a representative but bounded collaboration design space. Although the EVGen component of AssertMate can support different collaboration architectures, this study focuses on three representative mechanisms (i.e., majority voting, debate, and LLM-as-a-Judge) under the same perspective setting. These mechanisms represent direct consensus, iterative interaction, and expert-based selection, respectively, but they do not imply that the selected architecture is universally optimal. Future work will investigate adaptive perspective selection, dynamic agent allocation, and more advanced agentic collaboration strategies.

\subsection{Extensibility of Actual-Value Construction}

The current \AVCAgent{} rules (i.e., the return values and public getters) are designed as a practical approximation rather than an exhaustive semantic model of all possible observation points. In our Defects4J benchmark, these rules cover 86.5\% of the assertion targets appearing in developer-written tests. This result suggests that return values and object states exposed through public getters represent common and practically useful sources of test oracles. More importantly, these observation points can be reliably identified and reconstructed, enabling the generation of executable assertions while avoiding the uncertainty introduced by arbitrary method invocation.
The remaining 13.5\% of assertion targets mainly involve parameterized method invocations used directly as actual values, such as \texttt{assertEquals(expected, obj.computeSomething(param))}. Supporting such cases requires more than identifying observable values; it introduces a parameter-synthesis problem, where the framework must infer valid argument values, construct corresponding declarations and assignments, and ensure that the generated invocation is both type-correct and semantically meaningful in the test context. Therefore, we intentionally exclude these cases from the current design rather than introducing broad invocation rules that may generate uncompilable or unreliable assertions.

Nevertheless, extending \AVCAgent{} to support richer observation points is feasible. Potential directions include combining static analysis, constant propagation, and symbolic execution to infer valid arguments; mining project-specific test patterns to identify common invocation strategies; and leveraging LLM-based argument generation guided by method signatures and test context. These extensions are complementary to the current framework rather than fundamental changes to its core design. The existing rule-guided observation-point construction already provides broad practical coverage, while future extensions can further improve coverage for more complex assertion scenarios.

\subsection{Threats to Validity}

\noindent
\textit{External Threats}.
Our evaluation is based on Defects4J, a widely adopted benchmark comprising real-world Java defects. However, its limited dataset size and Java-specific characteristics may limit generalizability to other programming languages. To mitigate this, we selected 2,203 EvoSuite-generated FM--TP pairs when validating the integration of \TECH{} with EvoSuite, substantially expanding the experimental scale. 
Nevertheless, AssertMate is not inherently restricted to Java. 
The language-dependent components are mainly confined to ActVCon, which requires adapting program analysis and assertion mapping rules to different languages and testing frameworks, while the core LLM-based components remain language-independent.
For example, extending AssertMate to Python or C++ only requires replacing Java-specific analysis with corresponding mechanisms for attribute/member analysis and adapting assertion templates to frameworks such as pytest or GoogleTest. In future work, we plan to extend our evaluation to larger, multi-language datasets to validate generalizability further.

\smallskip
\noindent
\textit{Internal Threats.}
One internal threat stems from the inherent randomness of LLM outputs. 
Although we fixed the temperature setting, stochastic generation may still affect assertion reproducibility. 
To address this, we generated five responses per query and applied a probability-based re-ranking strategy to select the most reliable output, thereby mitigating variability.
Another potential threat is contamination in the backbone LLM. 
DeepSeek-Coder is trained on large-scale open-source code, while its pretraining corpus is not publicly available; therefore, we cannot fully verify whether Defects4J-related projects are included in its training data or perform strict decontamination. 
However, \TECH{} does not directly prompt the LLM to reproduce existing developer-written assertions. Instead, ActVCon first derives assertion targets through static analysis and type-aware rules, and the LLM subsequently predicts expected values for these constructed actual values. The generated assertions are further validated through compilation and execution. Therefore, even if related Defects4J code was encountered during pretraining, successful assertion generation requires context-specific oracle inference rather than simple memorization of existing assertions. While this design reduces the risk of contamination-driven performance gains, we cannot completely eliminate this threat.

\smallskip
\noindent
\textit{Construction Threats.}
One major construct validity threat lies in the evaluation of the pass rate and bug detection rate. 
Since we focus on the unit test scenario, the correctness of the focal method is often unknown, making it difficult to distinguish between true assertion failures and false alarms. To address this challenge, we adopted two mitigation strategies: (1) for evaluating \TECH{}’s effectiveness, we used human-written bug-triggering test cases from Defects4J, which are expected to pass on the fixed version and fail on the buggy version, thus serving as reliable ground truth; and (2) in RQ2, we used the number of killed mutants as an alternative evaluation metric, offering a more objective measure of assertion effectiveness in fault detection.

\section{Conclusion}
\label{conclusion}
We proposed \TECH{}, a novel agent-based assertion generation framework to facilitate generating effective assertions for automated testing. 
\TECH{} offers three key advantages: (1) it integrates static program analysis to generate actual values, effectively mitigating LLM hallucinations; (2) it leverages complementary prompting strategies for richer context understanding; and (3) it employs a tailored multi-agent collaboration mechanism to determine accurate expected values. Evaluation results demonstrate that \TECH{} significantly outperforms existing state-of-the-art techniques and validate the effectiveness of its core components.

This work opens several promising avenues for future research. One direction is to improve the efficiency of \TECH{} and extend its applicability beyond Java, broadening its impact across languages. 
Another is to explore more advanced collaboration mechanisms, such as dynamic task allocation, to better leverage the reasoning capabilities of large language models.
We also plan to address hallucinations in expected value generation and component coordination by incorporating techniques like knowledge distillation to improve consistency and reliability. Furthermore, evaluating \TECH{} in multi-assertion generation scenarios presents another valuable direction. While AssertMate already supports multi-assertion generation through \AVCAgent{}’s ability to synthesize multiple actual values for the same test prefix, the challenge of determining ``where to assert'', especially when multiple test scenarios are embedded within a single test case, remains underexplored across existing approaches, including ours.
Addressing this challenge is critical for scaling assertion generation from isolated unit behaviors to more realistic test cases.

\section*{Acknowledgment}
This work was supported by the National Key Research and Development Program of China (Grant No. 2024YFB4506300), the National Natural Science Foundation of China (Grant Nos. 62322208, 12411530122, 62232001).

\bibliographystyle{IEEEtran}
\bibliography{ref}

\end{document}